# Correlation between Microstructural and Magnetic Properties of Epitaxial YIG Films by Pulsed Laser Deposition

José Diogo Costa,* Niels Claessens, Giacomo Talmelli, Davide Tierno, Farah Amar, Thibaut Devolder, Matthijn Dekkers, Johan Swerts, Sean R. C. McMitchell, Florin Ciubotaru, and Christoph Adelmann*

**ABSTRACT:** In this study, we investigate the relationships among film growth conditions, crystalline microstructure, and magnetic properties of epitaxial yttrium iron garnet ($Y_3Fe_5O_{12}$, YIG) thin films, deposited on gallium gadolinium garnet ($Ga_3Gd_5O_{12}$, GGG). A direct correlation was observed between the residual epitaxial strain, bulk magnetic properties like effective magnetization and magnetic damping, and the performance of spin-wave transmission devices based on these films. This correlation offers a pathway for a simplified, rapid assessment of YIG film quality, avoiding the need for complex time-consuming characterization techniques. In addition, we report a comprehensive investigation into the influence of pulsed-laser deposition parameters, including deposition temperature, pressure, laser fluence, frequency, and annealing conditions. Through systematic deposition optimization, state-of-the-art YIG films exhibiting ultralow magnetic damping could be obtained, which is critical for high-performance spintronic applications.

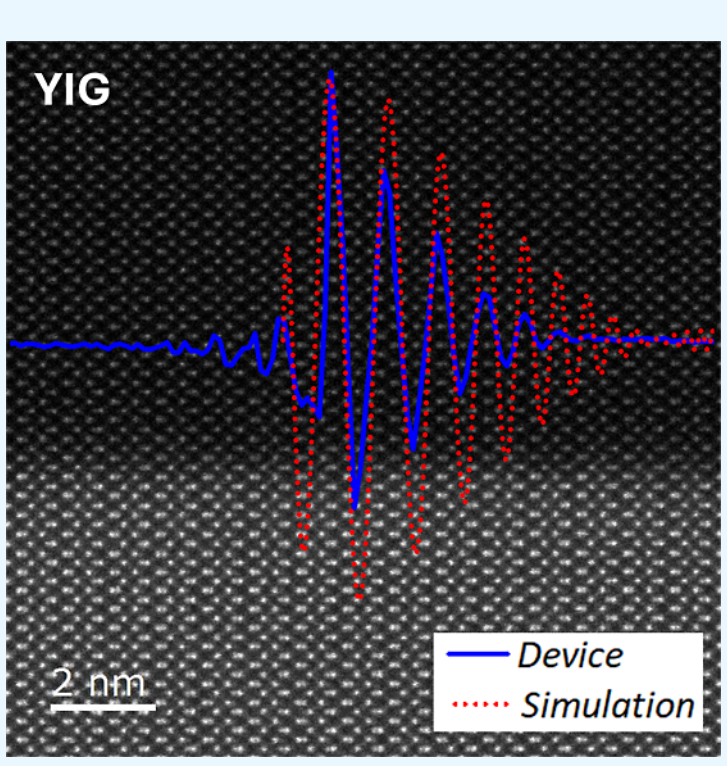

## 1. INTRODUCTION

Yttrium iron garnet ($Y_3Fe_5O_{12}$, YIG), with its exceptional combination of ultralow magnetic damping, high Curie temperature (560 K), large magnetic permeability, and optical transparency in the infrared range, has garnered much interest across various technological domains, in particular, for applications in microwave and spintronic technologies.[1−4] A pivotal factor in realizing the full potential of YIG is, however, the fabrication of high-quality epitaxial thin films. Epitaxial growth is crucial for attaining the desired magnetic properties, that are not observed in polycrystalline or amorphous films. Including some of the lowest magnetic damping of any known magnetic material leads to optimal device performance.

Epitaxial thin films of YIG are commonly grown on gadolinium gallium garnet ($Ga_3Gd_5O_{12}$, GGG) substrates owing to the excellent lattice match and the favorable thermal, mechanical, and microwave properties of GGG. Several techniques have been employed for the epitaxial growth of YIG, with liquid phase epitaxy (LPE), sputtering, and pulsed laser deposition (PLD) being the most prominent. LPE facilitates the deposition of high-quality single-crystal films with exceptionally low magnetic damping ($\alpha < 10^{-4}$).[5,6] However, depositing films thinner than a few hundred nanometers by LPE presents significant challenges.[5] By contrast, RF magnetron sputtering, while widely utilized for depositing dielectric films, often results in YIG films exhibiting O vacancies, leading to an increase in magnetic damping ($\alpha \approx 1 \times 10^{-3}$).[7−10]

In contrast, PLD offers exceptional control over film stoichiometry and epitaxy, rendering it highly suitable for depositing complex oxide materials. Notably, ultralow magnetic damping values ($\alpha \approx 10^{-4}$) have been reported for PLD-grown films with thicknesses as low as 20 nm.[11−20] Several PLD deposition parameters, including deposition temperature, pressure, laser fluence, and frequency, as well as annealing conditions, can be systematically adjusted to optimize film deposition. These parameters exert distinct and significant influences on the underlying film formation mechanisms.[21,22]

For instance, the O partial pressure exerts a significant influence on plume kinetics, enabling the fine-tuning of compositional variations between the target and the deposited film.[23−25] However, the optimal deposition pressure for achieving the highest film quality remains a subject of ongoing debate. While some studies have reported beneficial effects of higher deposition pressures (up to 0.67 mbar),[26] the majority of studies suggest a lower pressure regime (0.01−0.05 mbar) for optimal film growth.[11,23,27] Laser fluence significantly influences the shape and characteristics of the ablated material plume, impacting both the deposition rate and film composition. On the other hand, the frequency of laser pulses determines the time interval available for deposited material to find stable sites before the arrival of subsequent ablated adatoms, affecting adatom thermalization, diffusion, and ultimately the growth mode. However, the combined effects of laser fluence and frequency on the resulting film properties are often under-explored in the literature.[26]

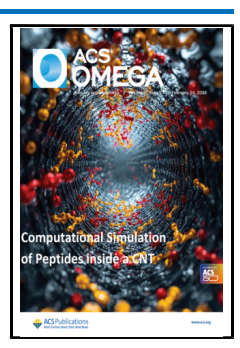

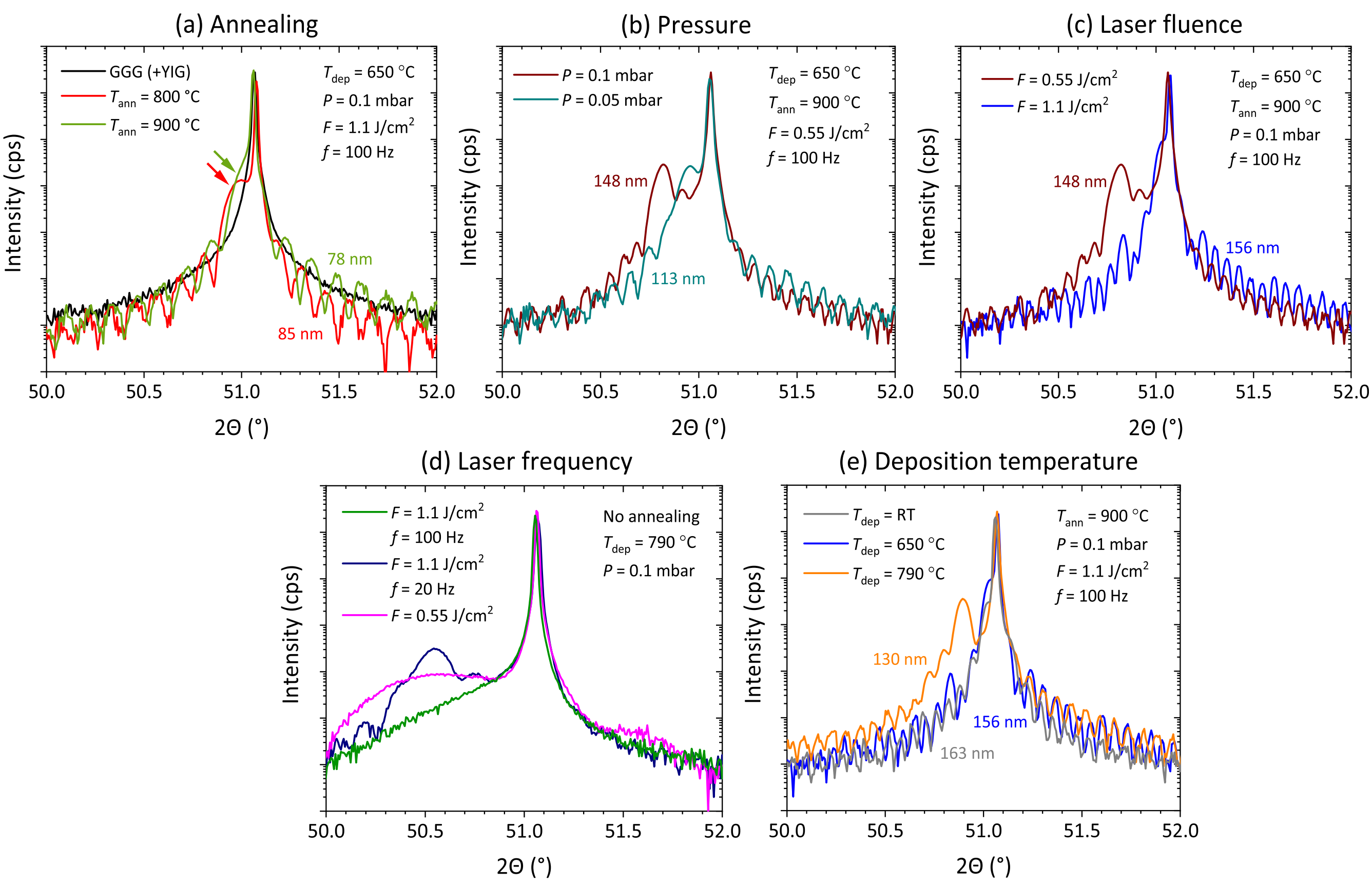

**Figure 1.** Structural characterization of epitaxial YIG films grown on GGG substrates using high-resolution XRD. (a) Reference XRD spectrum of the bare GGG substrate (black), compared with YIG films annealed at 800 °C (red) and 900 °C (green). (b) Influence of $O_2$ pressure during deposition: films deposited at 0.05 mbar (green) and 0.1 mbar (brown), both annealed at 900 °C. (c) Effect of laser fluence on film quality: spectra for YIG films deposited at 0.55 J/cm$^2$ (brown) and 1.1 J/cm$^2$ (blue), followed by annealing at 900 °C. (d) Impact of laser frequency and deposition rate: comparison of films deposited with a fluence of 1.1 J/cm$^2$ at 100 Hz (green), 1.1 J/cm$^2$ at 20 Hz (dark blue), and 0.55 J/cm$^2$ at 20 Hz (pink), not annealed. (e) Effect of deposition temperature: XRD spectra of YIG films deposited at RT (gray), 650 °C (blue), and 790 °C (orange), followed by annealing at 900 °C.

Growth temperature plays a crucial role by affecting the surface mobility of impinging adatoms and thus the crystal structure and atomic arrangement within the growing film. Reported PLD deposition temperatures for YIG range widely from room temperature (RT) to 850 °C. While RT deposition offers significant advantages from a processing perspective, e.g. by enabling lift-off techniques for subsequent patterning,[28,29] the majority of high-quality YIG film deposition processes still operate at elevated growth temperatures, typically around 650 °C with some variations reported.[14,19,20,30,31] Even in cases of high-temperature deposition, postdeposition annealing is frequently required to optimize film properties. Annealing in vacuum can lead to O out-diffusion and degrade film quality.[32] Therefore, annealing in an $O_2$ atmosphere is generally preferred to ensure optimal film properties.[23]

Hence, there is still a need for a comprehensive understanding of the roles that these parameters play in YIG growth. Furthermore, establishing a robust correlation between crystallographic characteristics and material—and ultimately device—performance would significantly expedite the optimization of YIG-based spintronic devices. In this study, we demonstrate a direct correlation between the crystal structure and magnetic properties of PLD deposited YIG films, enabling a simplified assessment of magnetic film properties prior to device fabrication. We also elucidate the relationship between the deposition conditions and film properties.

As a result, state-of-the-art YIG films with ultralow magnetic damping ($\alpha < 3 \times 10^{-4}$) were achieved under optimized conditions. The optimal deposition conditions, validated by spin wave simulations and device level measurements, were determined to be 650 °C for deposition temperature, 1.1 J/cm$^2$ for laser fluence, 0.04 mbar for $O_2$ pressure, in combination with postdeposition annealing at 900 °C. Furthermore, we also report a strong dependence of YIG properties on film thickness. Notably, as the film thickness increases, the YIG film exhibits a transition from an expanded to a compressed state relative to the lattice parameter of the GGG substrate. This comprehensive understanding of the film growth mechanisms is pivotal for the development of optimized spintronic devices.

## 2. EXPERIMENTAL DETAILS

All YIG depositions were carried out on (111)-oriented GGG substrates (1″ diameter, CrysTec) in a Solmates SIP-800 PLD system.[33] Samples were deposited under a range of deposition conditions, varying systematically temperature (up to 790 °C), $O_2$ pressure, laser fluence, and laser frequency. Subsequently, the films were annealed postdeposition at various temperatures under an $O_2$ pressure of $1.8 \times 10^{-2}$ mbar for 1 h, using a temperature ramp of 10 K/min for both heating and cooling cycles.

The structural, morphological, and magnetic properties of the YIG films were characterized using X-ray diffraction (XRD, Panalytical X'Pert Pro), scanning transmission electron microscopy (STEM), ferromagnetic resonance (FMR), Rutherford backscattering spectrometry (RBS), and elastic recoil detection (ERD) analysis (ERD). STEM images were obtained using a Tecnai F30 ST microscope operated at 300 kV. The samples were prepared by first capping them with a thin spin-on carbon (SOC) layer and then cutting 50 nm-thick lamellae using a Helios 450HP focused ion beam instrument. The RBS and ERD instrumentation is described elsewhere.[34] The film thickness was determined from the Laue oscillations in the XRD spectra and was subsequently confirmed through STEM analysis. The FMR measurements were performed using an Anritsu MS4645B vector network analyzer (VNA).

Device fabrication was carried out using a combination of hard mask definition and YIG wet etching. The hard mask (30 nm $SiO_2$) was patterned through $CF_4$ reactive ion etching, followed by wet etching of the YIG layer using phosphoric acid ($H_3PO_4$ 85%) at 130 °C. SOC was employed for planarization, followed by an additional deposition of 30 nm $SiO_2$ to enhance adhesion. Au antennas (200 nm thick, 5 $\mu$m wide) were subsequently defined using a lift-off process.[35]

## 3. RESULTS AND DISCUSSION

### 3.1. Crystalline Structure

The crystallographic structure of the YIG films was investigated by using XRD analysis, with a focus on comparing films of similar thicknesses. YIG, a ferrimagnetic garnet with cubic lattice structure, exhibits epitaxial growth on isostructural GGG substrates with low defect densities due to the small lattice mismatch of $\Delta a/a \approx 6 \times 10^{-4}$.[36] The (111)-oriented GGG substrate displays a well-defined (444) diffraction peak at $2\theta = 51.1°$ in the $2\theta-\omega$ scans [Figure 1a; black solid line]. Following YIG deposition at 650 °C, the XRD spectrum remained unchanged, indicating that the YIG film was X-ray amorphous. However, postannealing resulted in the emergence of a shoulder at low $2\theta$ in combination with distinct Laue oscillations in the diffraction pattern, which signifies the formation of a coherent crystalline YIG film.[37] These fringes were observed for both 800 °C (red solid line) and 900 °C (green solid line) postdeposition annealing temperatures in YIG films with comparable thicknesses (~80 nm). The shoulder could be attributed to the YIG (444) reflection with an out-of-plane lattice parameter slightly larger than those of bulk YIG and GGG.

A notable observation was the shift of the YIG (444) shoulder [indicated by arrows in Figure 1a] toward the main GGG (444) peak upon increasing the annealing temperature to 900 °C. This shift suggests a weaker rhombohedral distortion of the YIG film—and thus lower residual strain—at higher annealing temperatures. While 800 °C annealing yielded already high-quality epitaxial films, elevating the annealing temperature to 900 °C thus further enhanced the epitaxial film quality. Consequently, 900 °C was selected as the standard annealing temperature for the remainder of this study.

Although postdeposition annealing is crucial for achieving high-quality epitaxial films, further investigations have revealed a substantial influence of various PLD deposition parameters on the properties of the annealed YIG films. For instance, Figure 1b illustrates the impact of the PLD deposition pressure $P$, by contrasting the high-resolution XRD spectra of two films grown under identical conditions except for the $O_2$ pressure: 0.05 mbar (green line) and 0.1 mbar (brown line). We note that both films were subjected to identical postdeposition conditions.

It is well established that the pressure of the $O_2$ during deposition has a profound impact on the PLD growth mode, modifying both the ablation plume profile and the O content within the growing film. Although some studies have reported advantages of higher deposition pressures up to 0.67 mbar for YIG PLD,[26] several studies in the literature suggest that lower pressures on the order of 0.01−0.05 mbar typically yield superior film quality.[11,23,27] The data in Figure 1b indicate that the pressure of the $O_2$ during PLD significantly affects the rhombohedral distortion and the residual strain in the films, as evidenced by the shift of the YIG (444) reflection. Note that the shift is in the opposite direction, as expected for differences in the O content for different $O_2$ partial pressures. This emphasizes that lower pressures are beneficial for enhancing film properties, a factor that becomes increasingly critical for thicker films, as will be demonstrated in the subsequent section.

A second critical deposition parameter in the PLD of YIG is the laser fluence $F$ applied to the target. Laser fluence not only influences the deposition rate but also significantly impacts the profile of the ablated plume. Deposition occurring at the visible plume edge, i.e., closer to the substrate within the plume, generally results in more stoichiometric growth.[38] Outside the central region of the ablation plume, heavier elements experience increased scattering and reduced forward momentum in the ambient oxygen, leading to preferential loss and deviations from stoichiometric transfer.

Figure 1c presents a comparison of XRD spectra from YIG films deposited under identical conditions except for the laser fluence: 1.1 J/cm$^2$ (blue curve) and 0.55 J/cm$^2$ (brown curve). A significantly larger rhombohedral distortion (i.e., larger out-of-plane lattice parameter) can be deduced from the YIG (444) peak position for the lower fluence. This indicates that higher laser fluence promotes a more stoichiometric transfer from the target to the film, likely due to deposition occurring closer to the visible plume edge, despite the higher deposition rate. Furthermore, a combined analysis of RBS and ERD revealed subtle variations in the film stoichiometry. The calculated film compositions were determined to be $Y_{3.7 \pm 0.07}Fe_{4.9 \pm 0.1}O_{11.3 \pm 0.2}$ for high fluence and $Y_{3.7 \pm 0.07}Fe_{3.8 \pm 0.08}O_{12.5 \pm 0.2}$ for low fluence, with the stoichiometric composition being $Y_3Fe_5O_{12}$. These findings indicate that more stoichiometric iron transfer can be achieved with a higher laser fluence. While different laser energies can also potentially influence the growth mode (e.g., columnar growth), AFM analysis revealed a consistent surface roughness of 0.2 nm for both studied laser energies, suggesting a similar growth mode (data not shown). We note that this low surface roughness is crucial for minimizing magnetic damping within the YIG films.[19]

Although counterintuitive, postdeposition annealing is crucial for achieving epitaxial film growth despite the already rather elevated deposition temperature. Film growth is characterized by dynamic surface processes involving continuous adatom diffusion and nucleation, whereas postdeposition annealing predominantly induces defect diffusion and atomic rearrangement within the film bulk. Thus, the deposition temperature alone does not fully replicate effects of postdeposition annealing.

Eliminating the need for postdeposition annealing is a significant advantage from a fabrication perspective. Epitaxial films could potentially be obtained directly, without annealing, by employing significantly reduced deposition rates, which would increase the stabilization time of the adatoms. Both laser fluence and frequency influence the deposition rate, but their effects differ fundamentally. Laser fluence directly modifies the ablation plume, affecting the film stoichiometry. Conversely, the laser frequency modulates the temporal interval between successive plume pulses. At lower frequencies, adatoms have a longer time for thermalization and rearrangement before the arrival of the next pulse, potentially facilitating improved epitaxial growth. However, strongly reduced deposition rates severely limit process throughput. Consequently, a balance must be found between deposition rate and postdeposition annealing to achieve optimal film quality and practical processing times.

This is illustrated in Figure 1d, which presents the XRD spectrum of an as-deposited YIG film grown at the maximum temperature of the PLD system (790 °C) by using a fluence of 1.1 J/cm$^2$ and a frequency of 100 Hz, the standard frequency employed in this study. Under these conditions, only the GGG (444) substrate peak is discernible in the XRD spectrum, indicative of an amorphous YIG film. Reducing the laser pulse frequency to 20 Hz (dark blue line) resulted indeed in the emergence of weak crystallinity within the as-deposited film. However, further decreasing the deposition rate by reducing the laser fluence (pink line) appeared to diminish again the degree of crystallinity compared to the 20 Hz case. This observation aligns with the findings presented in Figure 1c, where higher laser fluences were shown to promote an improved epitaxy. Despite exhibiting some degree of crystallinity, the epitaxial quality of these as-deposited films remained however low. Consequently, achieving high-quality YIG films without postdeposition annealing presents a significant challenge and is not replicated by slower deposition. Although prior substrate thermal treatment has shown some promise to improve this issue,[16,30] it does not offer a substantial processing advantage compared to postdeposition annealing.

While the deposition temperature does not directly influence the stoichiometry or shape of the plasma plume, it significantly impacts the surface diffusion of impinging adatoms, leading to variations in atomic arrangements within the growing film. This ultimately results in differences in the crystallographic properties observed after postdeposition annealing. Figure 1e illustrates XRD spectra of YIG films deposited at different temperatures [790 °C, 650 °C, and RT] following identical postdeposition annealing at 900 °C. Epitaxial films were obtained for all deposition temperatures, albeit with distinct crystallographic characteristics. In particular, larger rhombohedral lattice distortion was observed for films deposited at 790 °C.

One crucial factor contributing to the exceptional properties of YIG films grown on GGG substrates is the close lattice match between their respective lattice constants ($a_{YIG}$ = 12.376 ± 0.001 Å, $a_{GGG}$ = 12.383 ± 0.002 Å; $\Delta a/a \approx 6 \times 10^{-4}$). However, it is possible that the differential thermal expansion coefficients of YIG and GGG contribute to a more pronounced lattice mismatch at the highest deposition temperature, resulting in increased residual strain within the film upon cooling.[39] In contrast, negligible crystallographic differences were observed between films deposited at RT and 650 °C. However, as will be demonstrated later, the FMR signal intensity for the RT-deposited film is notably lower, suggesting incomplete crystallization within this film.

## 3.2. STEM

A key finding of this study is the apparent discrepancy between the requirement for high-temperature annealing (900 °C) and the beneficial effect of intermediate deposition temperatures (optimal at 650 °C). To elucidate the underlying mechanisms, STEM analysis was conducted on films deposited at 650 and 790 °C following postdeposition annealing. For films deposited at 650 °C [Figure 2a], imaging by STEM revealed a remarkably sharp interface between the YIG film and the GGG substrate, characterized by a perfect lattice match and an absence of interfacial dislocations. Furthermore, these films exhibited a near-perfect crystalline structure with minimal defects [Figure 2b].

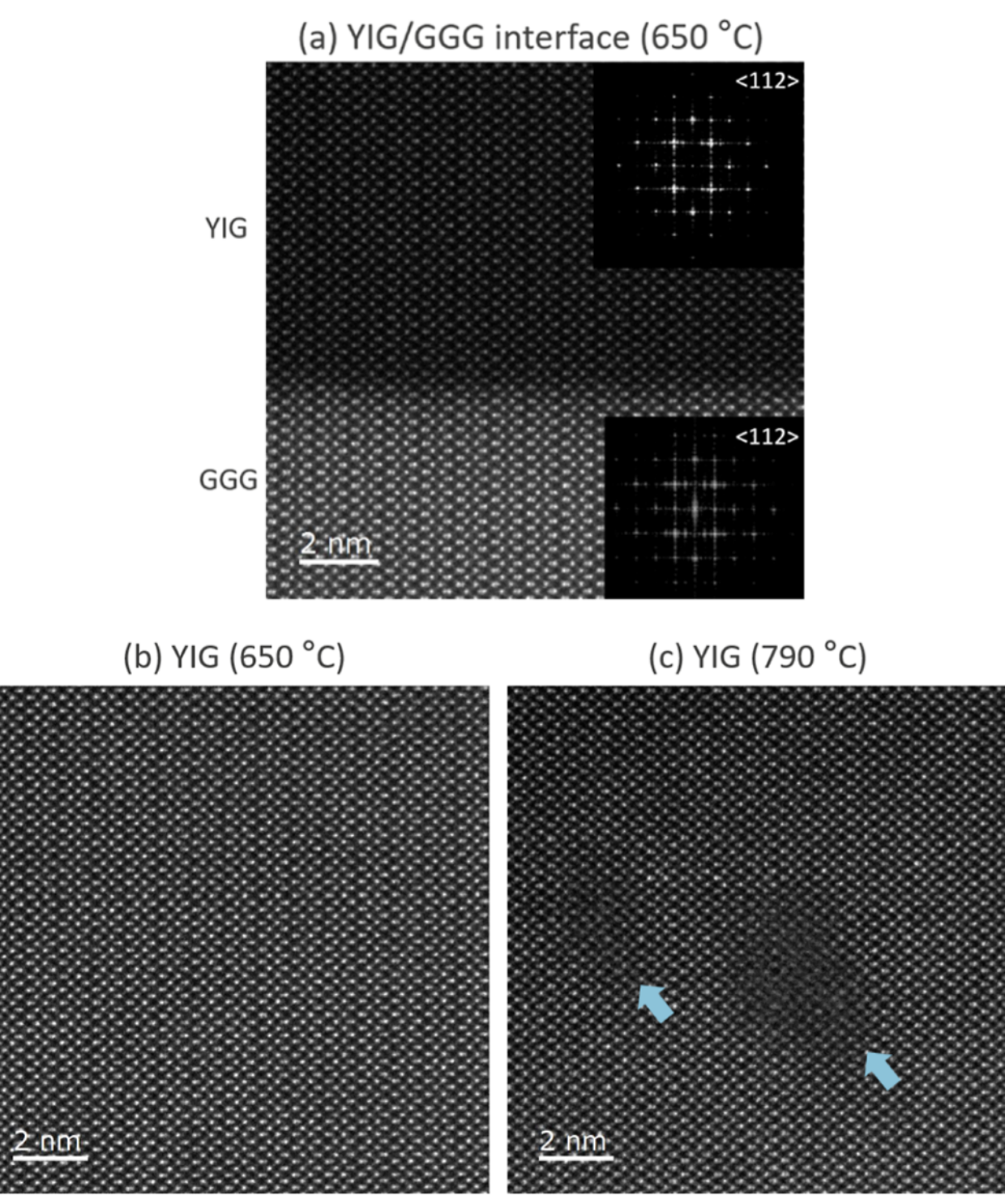

**Figure 2.** High-resolution STEM imaging of YIG films on GGG substrates. (a) Cross-sectional STEM image of the YIG/GGG interface for a deposition temperature of 650 °C, with insets showing selected area electron diffraction patterns for the YIG film and the GGG substrate, confirming epitaxial growth. (b,c) High-magnification STEM image of the YIG film deposited (b) at 650 °C and (c) at 790 °C. The blue arrows mark the presence of defects in the YIG film deposited at higher temperature.

In contrast, films deposited at 790 °C contained localized defects within the YIG lattice [Figure 2c, blue arrows]. These defects, distinct from (threading) dislocations, do not extend to the interface and manifest as elongated regions with varying contrast. Combined atomic bright-field and cross-sectional STEM analysis suggests that these regions may represent areas with a higher concentration of light elements (possibly O) and potentially exhibit local variations in crystallographic structure and/or amorphization. The reduced intensity of the atomic columns in bright-field imaging, compared to *Z*-contrast imaging, further supports the presence of local crystallographic variations (see Supporting Information). These localized regions likely introduce strain within the film, contributing to the larger rhombohedral distortion observed in the XRD spectra [Figure 1e]. These findings clearly demonstrate that despite the high-temperature annealing process lower deposition temperatures (i.e., 650 °C) yield significantly superior YIG film quality in terms of crystalline perfection and interfacial integrity.

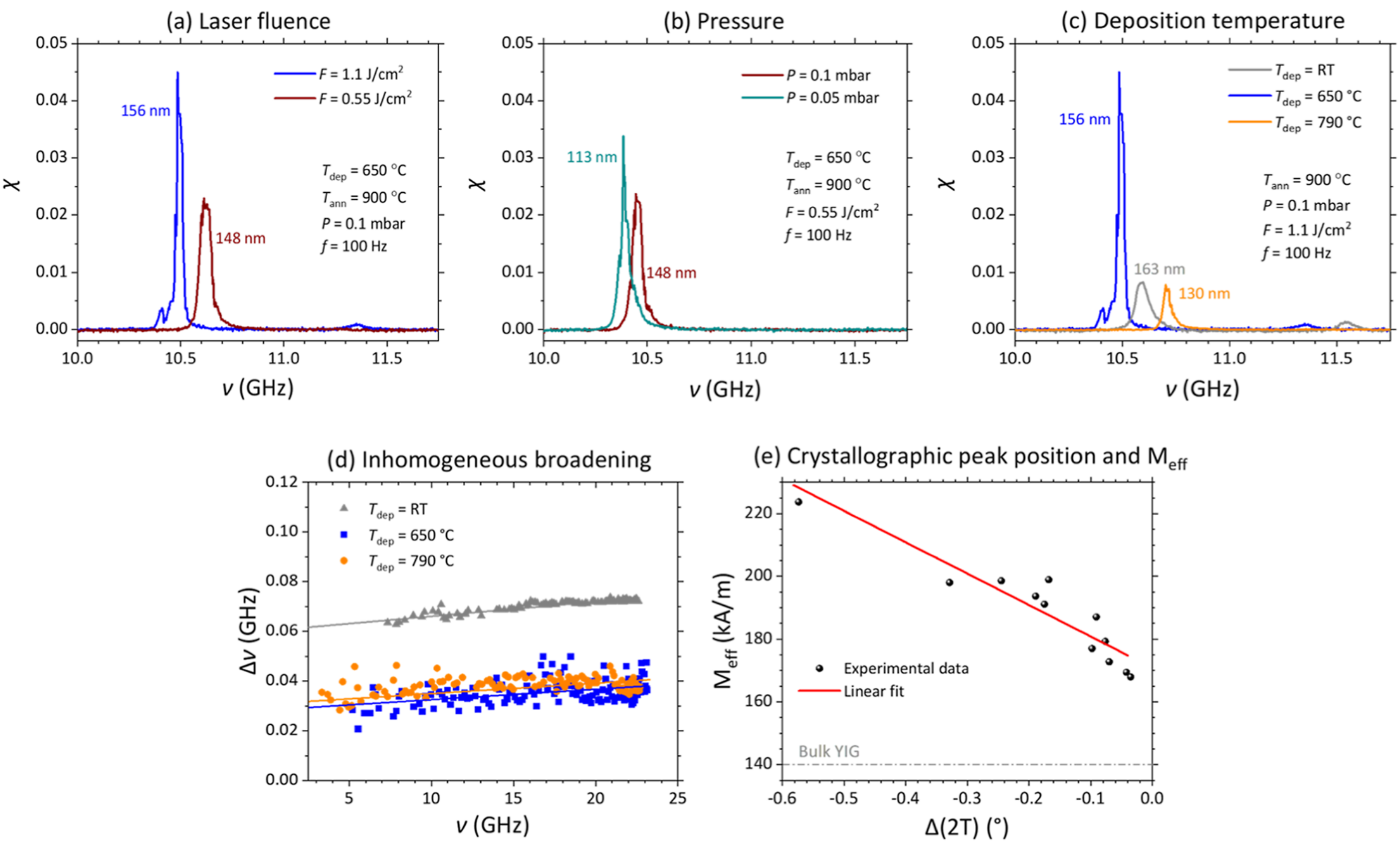

**Figure 3.** FMR characterization of YIG films deposited under various conditions. (a) Effect of laser fluence: FMR spectra for films deposited at 1.1 J/cm$^2$ (blue) and 0.55 J/cm$^2$ (brown). (b) Influence of $O_2$ pressure: FMR spectra for films deposited at 0.1 mbar (brown) and 0.05 mbar (green). (c) Deposition temperature dependence: FMR spectra for films deposited at RT (gray), 650 °C (blue), and 790 °C (orange). (d) Frequency-dependent FMR line width for films deposited at RT (gray triangles), 650 °C (blue squares), and 790 °C (orange circles). (e) Correlation between crystallinity and magnetic properties: Effective magnetization $M_{eff}$ as a function of the peak separation between YIG (444) and GGG (444) in the high-resolution XRD spectra.

### 3.3. Ferromagnetic Resonance

To correlate the magnetic properties with the previously established structural characteristics, VNA-FMR measurements were performed to characterize the magnetic dynamics of the YIG films. The measured absorption spectra were fitted to a Lorentzian curve, which enables the extraction of key magnetic parameters such as the effective magnetization $M_{eff}$ and magnetic Gilbert damping $\alpha$ from the resonance frequency and line width, respectively.[40,41]

The determination of ultralow Gilbert damping in state-of-the-art YIG films is inherently limited by the finite dimensions of the coplanar waveguide (CPW) used for the FMR measurements. As reported in ref 42 and detailed in the Supporting Information, the finite size of the CPW leads to a broadened emission spectrum, resulting in an apparent line width that is independent of the intrinsic film damping for values below approximately $3 \times 10^{-4}$ (calculated for a 150 nm thick film at 6 GHz). While the FMR technique accurately resolves damping values characteristic of materials such as CoFeB or permalloy (NiFe) with $\alpha \approx 10^{-3}$, the resolution limit imposed by the CPW dimensions impedes the precise quantification of ultralow damping values characteristic of epitaxial YIG films with $\alpha \approx 10^{-4}$. To corroborate this limitation on the FMR damping resolution, measurements were performed by using three independent FMR setups. Furthermore, repeated measurements (up to five times) on the same ultralow damping samples yielded a range of damping values between $0.3 \times 10^{-4}$ and $2 \times 10^{-4}$, consistent with the estimated line width broadening due to the finite CPW dimensions ($\alpha < 3 \times 10^{-4}$). Based on these observations, a lower limit of $3 \times 10^{-4}$ was adopted for the magnetic damping in this study.

By the FMR measurements, we could unequivocally establish a strong correlation between the epitaxial quality of YIG films (cf. Figure 1) and their magnetic damping. Specifically, all samples exhibiting well-defined Laue fringes in XRD measurements demonstrated ultralow magnetic damping values ($\alpha < 3 \times 10^{-4}$), while those lacking such fringes showed negligible FMR signals. Furthermore, non-annealed samples deposited at a lower laser frequency ($f$ = 20 Hz), despite exhibiting some degree of crystallinity as evidenced in Figure 1d, exhibited significantly higher damping values of $\alpha = 4 \times 10^{-3}$ and $6 \times 10^{-3}$ for fluences of 1.1 and 0.55 J/cm$^2$, respectively. The lower damping observed at higher fluence corroborates the findings in Figure 1c, which demonstrate the beneficial effects of higher fluence on the film stoichiometry. This suggests that reducing the laser frequency can improve the film crystallinity up to a certain point, beyond which adatom thermalization becomes limiting, ultimately hindering further improvement while reducing the deposition rate. In contrast, the laser fluence exhibits an optimal value that facilitates stoichiometric deposition at the visible plume edge, as previously discussed.[38]

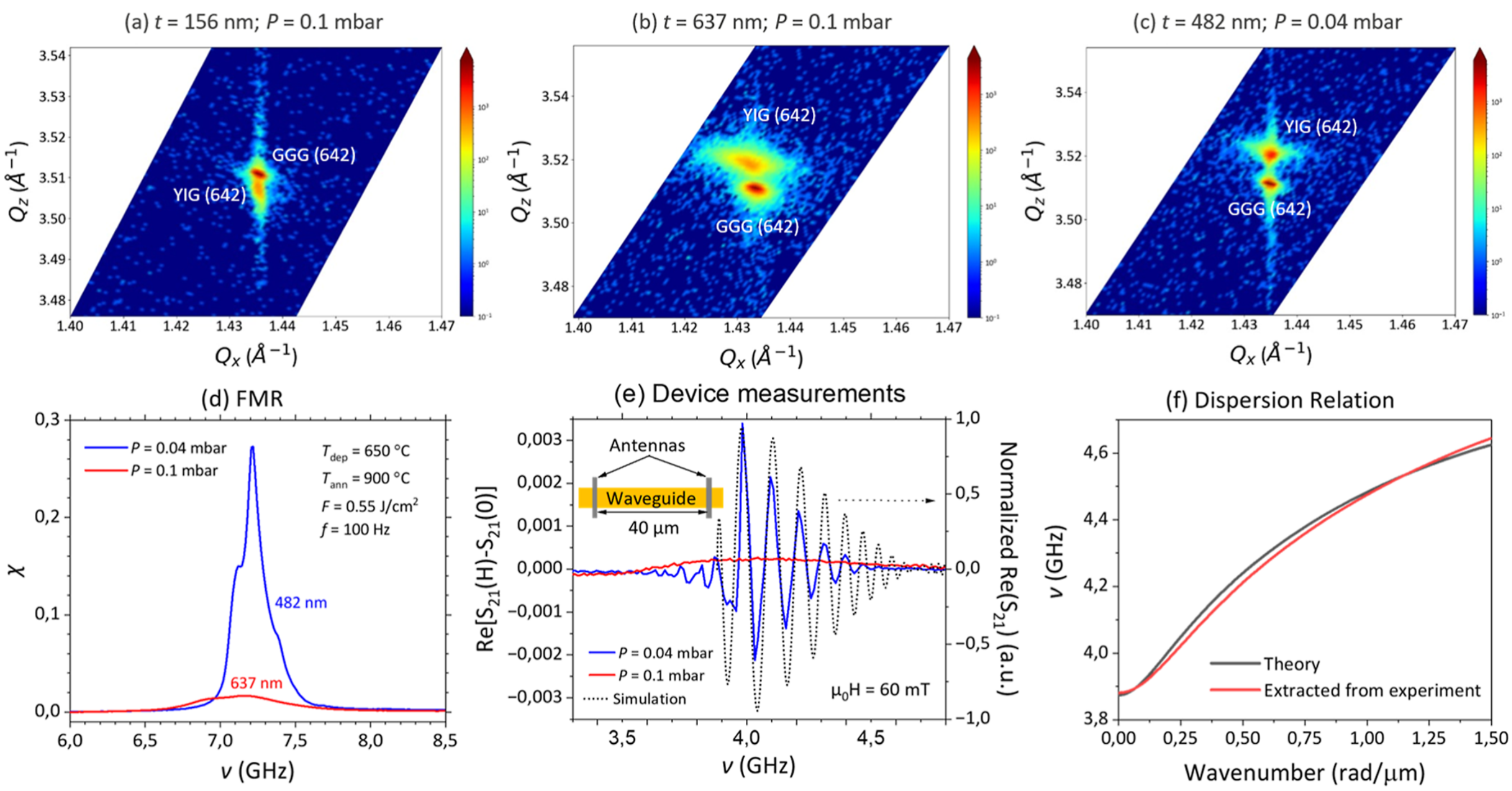

**Figure 4.** Material characterization and device-level measurements. RSMs centered around the YIG and GGG (642) reflections for samples with varying film thicknesses and oxygen pressures: (a) $t$ = 156 nm, $P$ = 0.1 mbar; (b) $t$ = 637 nm, $P$ = 0.1 mbar; (c) $t$ = 482 nm, $P$ = 0.04 mbar. All samples were deposited with the following optimized parameters: deposition temperature $T_{dep}$ = 650 °C, laser fluence $F$ = 1.1 J/cm$^2$, repetition rate $f$ = 100 Hz, and annealing temperature $T_{ann}$ = 900 °C. (d) FMR measurements and (e) all-electrical spin-wave transmission at the device level for the thicker YIG films deposited at $P$ = 0.04 mbar (blue) and $P$ = 0.1 mbar (red), and the simulated spin wave transmission (dashed black line). The inset in (e) illustrates the schematic of the fabricated device structure. (f) Dispersion relation calculated by iteratively fitting the model to the experimental measurements.

To further investigate the magnetic properties, we have also analyzed the FMR susceptibility, which is defined as

$$\chi = \frac{\ln S_{11}(H)}{\ln S_{11}^{\mathrm{ref}}} - 1 \tag{1}$$

where $S_{11}(H)$ is the microwave reflection $S$-parameter at a given magnetic bias field $H$, and $S_{11}^{\mathrm{ref}}$ is the signal at zero applied magnetic field. This quantity directly correlates with the spin-wave transmission signal measured at the device level, as demonstrated below. Additionally, for films with similar magnetic moments, the peak height can act as an indicator of the actual peak width since the integrated signals should be comparable. Therefore, higher signals should correspond to narrower bandwidths, indicative of the frequency independent line width broadening (inhomogeneous broadening).

Figure 3a−c present comparative FMR data for samples deposited under varying conditions, mirroring the approach employed in Figure 1. The FMR spectra were acquired within the 10−11 GHz frequency range under an out-of-plane magnetic field of 520 kA/m. A clear correlation emerges between the XRD analysis and the FMR signal intensity: closer YIG (444) and GGG (444) XRD peaks consistently correspond to stronger FMR signal. This observed correlation holds true across all investigated deposition parameters, including laser fluence [Figure 3a], O pressure [Figure 3b], and deposition temperature [Figure 3c].

It is important to note that the extracted damping values only reflect the frequency-dependent line width broadening (homogeneous broadening). In contrast, the magnon lifetime is intrinsically influenced by both homogeneous and inhomogeneous broadening mechanisms. Therefore, while low damping is a necessary condition for extended magnon lifetimes, it may not be a sufficient condition.

Figure 3d illustrates the line width (full width at half-maximum, FWHM) as a function of frequency $\nu$ for films deposited at different temperatures, corresponding the conditions in Figure 3c. Despite exhibiting similar intrinsic Gilbert damping values $\alpha$, as determined from the slope of line width vs frequency plots, the films displayed variations in inhomogeneous broadening (measured at $\nu$ = 0). Notably, the film deposited at RT exhibited a much higher degree of inhomogeneous broadening of around 3.4 Oe, compared to the films deposited at higher deposition temperature depicting lower values around 1.7 Oe. This suggests a shorter magnon lifetime compared with films deposited at higher temperatures. Contrary to some previous reports,[11,23] our findings indicate that deposition temperatures on the order of 650 °C yield superior film quality in terms of both low damping and minimized inhomogeneous broadening, ultimately leading to longer magnon lifetimes.

These results reveal a clear correlation between the crystallographic quality and/or stoichiometry of the YIG films and their FMR behavior. Specifically, the degree of rhombohedral distortion and residual strain, as evidenced by the separation between the YIG (444) and GGG (444) peaks in the XRD spectra (see Figure 1), emerges as a reliable indicator of the magnetic properties of the YIG films. While the precise quantification of ultralow damping values presents experimental challenges, as discussed above, and the comparison of FMR signal magnitudes is only practical for films of comparable thickness. On the other hand, the effective magnetization, which is given by the difference between saturation magnetization ($M_s$) and an out-of-plane anisotropy field ($H_k$), i.e., $M_{eff} = M_s - H_k$, can provide a valuable metric for assessing the magnetic quality of the films.

Figure 3e depicts the dependence of the effective magnetization $M_{eff}$ on the angular separation between the YIG (444) and GGG (444) peaks in the XRD spectra. A lower $M_{eff}$ value is indicative of both a smaller anisotropy field and an $M_s$ approaching the bulk value of YIG (~140 kA/m), suggesting properties that more closely resemble those of bulk YIG. The observed correlation establishes a quantitative link between the crystallographic quality of the YIG films and their magnetic properties. Both structural and magnetic characteristics are inherently shaped by factors such as stoichiometric variations and atomic arrangements, which result in the observed correlation.

The $M_{eff}$ values higher than $M_s$ for all conditions can be explained by the presence of an anisotropy field, that may originate from stoichiometric variations, magnetocrystalline or magnetoelastic contributions, and are consistent with other $M_{eff}$ values for PLD-grown YIG reported in the literature.[43] Vibrating sample magnetometry (VSM) measurements confirm that the out-of-plane (OOP) component increases with rising $M_{eff}$ (see Supporting Information, Figure S5), suggesting that the enhancement of $M_{eff}$ is associated with a more pronounced OOP component as film distortion increases. However, because of the low magnetic signal of the ferrimagnetic material and the diamagnetic contribution of the sample holder, the absolute magnetization could not be reliably extracted from the VSM measurements.

These findings can provide a valuable recipe for tuning the growth conditions of YIG films through the analysis of XRD data, without the need for microwave FMR measurements, and prior to spintronic device fabrication. It is noteworthy that previous literature reports have not consistently demonstrated a clear correlation between magnetization saturation and damping.[32] This discrepancy may be attributed to the inherent limitations in accurately quantifying ultralow damping values, as discussed earlier.

### 3.4. YIG-Based Spin-Wave Transmission Devices

Having established deposition conditions for high-quality epitaxial YIG growth ($T_{dep}$ = 650 °C, $F$ = 1.1 J/cm$^2$, $f$ = 100 Hz, $T_{ann}$ = 900 °C, and $P$ = 0.1 mbar), we proceeded to deposit thicker YIG films. This facilitates subsequent device-level characterization due to an enhanced magnetic signal. Figure 4a,b display reciprocal space maps (RSMs) acquired around the (642) reflections of YIG and GGG substrates for YIG films grown under these conditions with thicknesses of 156 and 637 nm, respectively. The RSMs corroborate the high-resolution XRD analysis in Figure 1, offering further insights into the film's crystallographic orientation.

A notable observation is that the YIG (642) peak exhibits a thickness-dependent shift in the reciprocal space, crossing the GGG substrate peak position. Initially, for the 156 nm-thick film, the slightly larger lattice constant of GGG results in compressive out-of-plane strain in the YIG film, consistent with the XRD results in Figure 1. However, as the YIG film thickness increases to 637 nm, the YIG lattice relaxes toward the equilibrium lattice constant, which is smaller than that of GGG out-of-plane. Consequently, the film transitions from a rhombohedrally distorted to a more relaxed cubic structure, albeit exhibiting increased broadness due to defect generation and larger out-of-plane and in-plane distortion. Considering the bulk lattice constants of YIG (12.376 Å) and GGG (12.383 Å), the observed relaxed YIG lattice parameter clearly indicates a lattice mismatch between the film and the substrate, possibly due to stoichiometric variation, leading to a larger relaxed film unit cell.

Based on the preceding analysis, a reduced $O_2$ pressure is preferred for improved epitaxial film quality. Figure 4c presents the RSM of a thicker YIG film ($t$ = 482 nm) deposited under the same conditions as those before, except with a lower $O_2$ pressure of 0.04 mbar. In contrast to the higher $O_2$ pressure results, this film still exhibits fully strained, pseudomorphic growth. However, the YIG (642) peak remains shifted to the opposite side of the GGG (642) peak compared with the thinner films. This observation underscores the significant impact of film thickness on the YIG lattice constant, leading to either a compressive or a tensile strain state. Moreover, the thinner film deposited at the nonoptimal pressure set point [156 nm; Figure 4a] still exhibits a YIG peak more closely matched with GGG than the thicker sample grown at the optimized set point [482 nm; Figure 4c]. This clearly demonstrates that film thickness is itself a factor affecting the properties of the PLD-grown YIG films.

While numerous studies on PLD have investigated YIG films with nm thicknesses,[11,14,16,19,20,24] the pronounced influence of film thickness on the resulting YIG properties has not been previously reported. Our findings demonstrate that PLD is particularly well-suited for the deposition of thin YIG films, including those in the submicrometer range, while LPE may be more advantageous for applications requiring thicker films.[35] However, by optimizing the PLD deposition parameters, it is possible to achieve high-quality YIG films with thicknesses exceeding the micrometer range.[31,44]

The crystallographic quality of the films profoundly influences their magnetic properties. Figure 4d presents FMR spectra of the two thick YIG films deposited at different $O_2$ pressures: 0.1 (637 nm; red line) and 0.04 mbar (482 nm; blue line). The pseudomorphic film deposited at the lower pressure exhibits a significantly higher and sharper FMR signal, despite its smaller thickness. However, the presence of multimodal precession around the main resonance peak suggests a more pronounced granular structure within the thicker film. This multimodal precession precluded a reliable Lorentzian fit to the FMR data, preventing accurate determination of the Gilbert damping parameter $\alpha$ and effective magnetization $M_{eff}$.

Both sets of deposition conditions, utilizing thicker YIG films, were employed for the fabrication of spin-wave transmission devices. The fabricated devices consisted of a 15 $\mu$m wide YIG spin-wave waveguide and two inductive antennas separated by 40 $\mu$m [inset of Figure 4e]. Two distinct signal propagation behaviors were observed for devices fabricated using YIG films grown under two different pressure conditions [Figure 4e]. The device fabricated using the YIG film grown at 0.04 mbar exhibited a substantially larger signal amplitude, consistent with the enhanced FMR response observed for this film. Furthermore, clear phase oscillations were observed in the signal transmitted through this device, a well-established hallmark of spin-wave propagation, indicating efficient device performance. In contrast, only incoherent precession was observed in the device fabricated using the YIG film grown at 0.1 mbar. To confirm that the observed effects are not solely due to film thickness, we deposited a film with half the number of laser pulses (~318 nm thick) at the highest pressure set point (0.1 mbar). This thinner sample also failed to exhibit coherent spin-wave propagation (not shown), corroborating that a lower pressure set point is required for functional spintronic device fabrication.

These findings demonstrate a strong correlation between the material properties, of the YIG films, including their crystallographic quality and magnetic properties, and the resulting device functionality. Coherent spin wave propagation was observed over distances as large as 115 $\mu$m (the maximum device distance) in the device fabricated by using the high-quality, low-pressure deposited YIG film. We note that despite the observation of multimodal FMR signals in the bulk material, a clear, apparently single-mode signal was measured in the spin-wave transmission devices.

In fact, we carried out an iterative fitting procedure in which the dispersion relation and spin-wave propagation characteristics were calculated analytically, compared with experimental data, and subsequently used to refine the model.[45,46] As shown in Figure 4f, the reconstructed dispersion relation agrees well with the analytical model and enables an accurate calculation of the group velocity. Using this velocity together with the antenna distribution in $k$-space, we extracted the spin-wave transmission, which closely reproduces the experimentally observed behavior [dashed black line in Figure 3e]. The theoretical dispersion converged to parameter values fully consistent with the measurements: a YIG thickness of 485 nm, a saturation magnetization of 153 kA/m, and an effective field of 74.5 mT, for a fixed film width of 15 $\mu$m and a $g$-factor of 2. This convergence suggests that the multimodal features observed in the FMR spectra likely originate from magnetic inhomogeneities on longer length scales. Such inhomogeneities have negligible impact on spin-wave propagation within the micrometer-scale devices, which therefore operate effectively in a single-mode regime at the device level.

## 4. CONCLUSIONS

In conclusion, we have demonstrated a strong correlation between the crystallographic YIG (444) peak position, readily assessable via high-resolution XRD, and the bulk magnetic properties and subsequent spin-wave propagation. These findings offer a promising approach for the rapid and simplified assessment of YIG film quality, potentially obviating the need for more complex and costly characterization techniques such as VNA ferromagnetic resonance.

Furthermore, this work provides a comprehensive overview of the optimized PLD parameters for the growth of state-of-the-art YIG films. These optimized conditions (fluence of 1.1 J/cm$^2$, deposition pressure of 0.04 mbar, deposition temperature of 650 °C, and annealing temperature of 900 °C) yield epitaxial films with minimal defects and exhibit ultralow magnetic damping ($\alpha < 3 \times 10^{-4}$).

This study also sheds light on the impact of film thickness on the properties of PLD-grown YIG films, highlighting the inherent scalability advantages of PLD for the fabrication of devices utilizing sub$\mu$m thick YIG films. The optimized deposition parameters, validated through device-level measurements and single-mode spin wave simulations, provide a robust foundation for the future development of high-performance YIG-based spintronic devices.

## ■ ASSOCIATED CONTENT

### Supporting Information

The Supporting Information is available free of charge at https://pubs.acs.org/doi/10.1021/acsomega.5c12736.

Annular bright-field and Z contrast STEM imaging of the defect region in the YIG film deposited at 790 °C; energy-dispersive X-ray spectroscopy (EDS) atomic concentration profiles for YIG films deposited at 650 °C and 790 °C; RBS and ERD measured spectra for the sample deposited at 650 °C; VNA ferromagnetic resonance (VNA-FMR) analysis; and VSM measurement in the in-plane and out-of-plane directions (PDF)

## ■ AUTHOR INFORMATION

### Corresponding Authors

**José Diogo Costa** – *IMEC, Leuven 3001, Belgium; Centro Singular de Investigación en Química Biolóxica e Materiais Moleculares (CIQUS), Universidade de Santiago de Compostela, Santiago de Compostela 15782, Spain; Departamento de Química-Física, Universidade de Santiago de Compostela, Santiago de Compostela 15782, Spain;* orcid.org/0000-0001-5092-4824; Email: josediogo.teixeira@usc.es

**Christoph Adelmann** – *IMEC, Leuven 3001, Belgium;* Email: christoph.adelmann@imec.be

### Authors

**Niels Claessens** – *IMEC, Leuven 3001, Belgium;* orcid.org/0000-0002-8863-9532

**Giacomo Talmelli** – *IMEC, Leuven 3001, Belgium*

**Davide Tierno** – *IMEC, Leuven 3001, Belgium;* orcid.org/0000-0003-4915-904X

**Farah Amar** – *Centre de Nanosciences et de Nanotechnologies, Université Paris-Saclay, CNRS, Palaiseau 91120, France;* Present Address: Keysight Technologies Inc. Santa Rosa, CA, US 95403

**Thibaut Devolder** – *Centre de Nanosciences et de Nanotechnologies, Université Paris-Saclay, CNRS, Palaiseau 91120, France*

**Matthijn Dekkers** – *SolMateS BV, Enschede 7521 PE, The Netherlands;* Present Address: Lam Research Corporation, Fremont, CA, US 94538–0000.

**Johan Swerts** – *IMEC, Leuven 3001, Belgium*

**Sean R. C. McMitchell** – *IMEC, Leuven 3001, Belgium;* orcid.org/0000-0002-9916-0973

**Florin Ciubotaru** – *IMEC, Leuven 3001, Belgium*

Complete contact information is available at: https://pubs.acs.org/10.1021/acsomega.5c12736

### Notes

The authors declare no competing financial interest.

## ■ ACKNOWLEDGMENTS

J.D.C. acknowledges financial support from the European Union MSCA-IF Neuromag under Grant Agreement No. 793346. F.C.'s and C.A.'s contributions have been funded in part by the European Union's Horizon 2020 research and innovation program within the FET-OPEN project CHIRON under grant agreement No. 801055. Funding for open access charge: Universidade de Santiago de Compostela/CISUG.

## ■ REFERENCES